\documentclass[12pt]{iopart}

\usepackage{iopams,graphicx,amssymb}

\newcommand{\ket}[1]{\left|#1\right>}

\begin{document}

\title[Indicators of wavefunction (de)localisation for avoided crossing]{Indicators of wavefunction (de)localisation for avoided crossing in a quadrupole quantum billiard}

\author{Kyu-Won Park$^{1, 2}$, Juman Kim$^2$, Jisung Seo$^2$, Songky Moon$^{3}$ and Kabgyun Jeong$^{1,*}$}
\address{${}^1$ Research Institute of Mathematics, Seoul National University, Seoul 08826, Korea}
\address{${}^2$ Department of Physics and Astronomy, Seoul National University, Seoul 08826, Korea}
\address{${}^3$ Faculty of Liberal Education, Seoul National University, Seoul 08826, Korea}
\ead{kgjeong6@snu.ac.kr}
\vspace{10pt}
\begin{indented}
\item[]June 2021
\end{indented}

\begin{abstract}
The relationship between wavefunction (de)localisation and avoided crossing in a quadrupole billiard is analysed  The following three-types of measures are employed for wavefunction (de)localisation: inverse participation ratio, inverse of R\'{e}nyi entropy, and root-mean-square (RMS) image contrast. All these measures exhibit minimal values at the centre of the avoided crossing, where the wavefunction is maximally delocalised. Our results indicate that these quantities can be sufficient for the indication of wavefunction (de)localisation.
\end{abstract}

%
\vspace{2pc}

\section{Introduction}
Avoided crossing is a phenomenon in which two eigenvalues of the Hamiltonian of a quantum system come closer when a system parameter is varied, but they repel each other~\cite{vNE29}. Since the introduction of quantum mechanics, avoided crossing has been fundamental because it manifests the presence of subtle interactions between states in the quantum system. It has been extensively studied in various physical systems such as nanowires ~\cite{FFS+07,PKD17}, crystalline solids~\cite{BCZ19}, topological magnons~\cite{TKBS19}, charged ions~\cite{BSB+16}, and quantum billiards~\cite{ZOHY10,SGWC13}. Particularly in the case of cavity resonators (related to avoided crossing), several intriguing physical properties of exceptional points~\cite{SKML+16,BDHM+14}, unidirectional emission~\cite{RLK09}, and dynamical tunnelling~\cite{SGLX14} have been explored. Avoided crossing is basically considered based on the behaviour of eigenvalues~\cite{vNE29}. However, in general, the behaviours of the wavefunctions also undergo certain changes from the influence of avoided crossing~\cite{TT92, ER17, RM11}. Here, we examine this issue in the framework of the wavefunction (de)localisation.

Localisation or delocalisation is a useful notion to probe spatial features and also determine essential physical properties. For example, conductance in condensed matter physics~\cite{OGH+15,HF07} is controlled by the degree of localisation; the localised and delocalised electron wavefunctions correspond to the insulators and conductors, respectively. Anderson localisation has been investigated for systems such as aperiodic super lattices~\cite{JVT19} and monolayer graphene~\cite{SF18}. In quantum billiard problems, wavefunction localisation is also an important issue for the scar, which refers to the wavefunction localisation on unstable periodic orbits in the semi-classical limit~\cite{H84}. Numerous scar phenomena have been observed in various chaotic systems such as molecular systems, atomic systems, and micro-cavities~\cite{ABB98,WDWG98,CC96}. Following these pioneering studies, the study of quasi-scars~\cite{W06,UWH08} (without any association with classical periodic orbits) as well as scar-like modes in integrable systems~\cite{LRR+04,RL11} were conducted. Furthermore, localisation associated with chaotic diffusion has also been studied~\cite{MLR+18,GCQ+20}.

In our previous studies~\cite{PMSK+18,JPK20}, we investigated the Shannon entropy and its relation to avoided crossing in the microcavity. We found that the Shannon entropies increase as the centre of the avoided crossing is approached, and are exchanged across the avoided crossing. The definition of Shannon entropy is a relevant measure of the average amount of information for a random variable with a specific probability distribution~\cite{S48}; however, the physical meaning of Shannon entropy in our system still remains unclear. Therefore, we elucidate it in the view point of the wavefunction (de)localisation in this work. For this purpose, we study the relation between the Shannon entropy to inverse participation ratio~\cite{EM00,MWA11}, R\'{e}nyi entropy~\cite{R61,D19} and the root-mean-square (RMS) image contrast~\cite{SP94,P90,P87}.

The remainder of this paper is organised as follows. In Sec.~\ref{Helm}, the avoided crossing of two interacting modes in a quadrupole billiard is introduced. In Sec.~\ref{inverse}, we study the inverse participation ratio. The R\'{e}nyi entropy is discussed in Sec.~\ref{Renyi}. The RMS contrast is studied in Sec.~\ref{rms}. In Sec.~\ref{comparison}, we compare the three measures based on the wavefunction delocalisation. Finally, we summarise our work in Sec.~\ref{conclusion}.

\section{Avoided Crossing of Two Interacting Modes} \label{Helm}

\begin{figure}[t!]
\centering
\includegraphics[width=8.5cm]{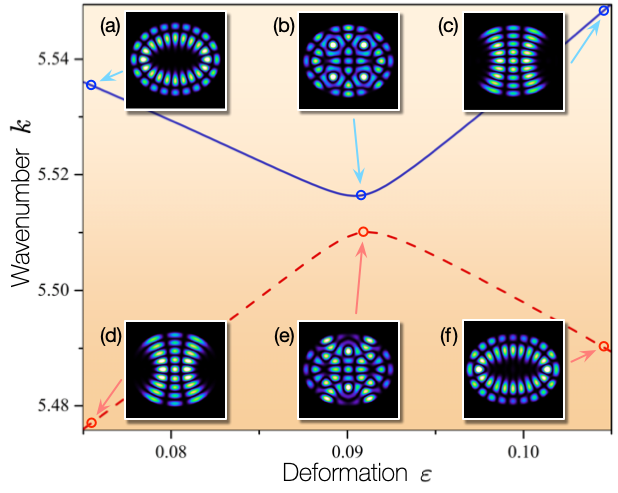}
\caption{Eigenvalues (or wavenumbers) of two interacting eigenmodes near the centre of the avoided crossing and the intensities of representative wavefunctions for a quadrupole billiard as a function of deformation parameter $\varepsilon$. The parameter $\varepsilon$ is restricted in $0.75\leq\varepsilon\leq 0.11$ at $n=3.3$. The value of solid blue curve and that of dashed red curve at $\varepsilon=0.75$ is approximately 5.5858 and 5.4423, respectively. The avoided crossing occurs at $\varepsilon\approx 0.091$. The wavefunctions corresponding to (a) and (d) are mixing together at (b) and (e) and then exhibit mode exchange at (c) and (f).}
\label{Fig.1}
\end{figure}

Now, we consider a closed \emph{quadrupole} cavity as a model for the chaotic billiard in order to study the avoided crossing; the non-linearity in a chaotic system leads to the avoided crossing~\cite{H99,F10}.  The geometric shape of the quadrupole is precisely described by $r(\phi)=\frac{1}{\sqrt{1+\varepsilon^{2}/2}}(1+\varepsilon\cos(2\phi))$, where $\varepsilon$ is the deformation parameter and $\phi$ is the angle in the polar coordinate. The pre-factor ${1}/{\sqrt{1+\varepsilon^{2}/2}}$ is adopted for an area-preserving condition. We obtain the eigenvalues and their wavefunctions by solving the Helmholtz equation $\nabla^{2}\psi+n^{2}k^{2}\psi=0$ equipped with the boundary element method~\cite{W03} for the transverse magnetic modes. We used the Dirichlet boundary conditions for closed (billiard) systems. Here, $k$ denotes the (vacuum) wavenumber, $n$ is the refractive index of the cavity whose value is $n=3.3$ in this paper and $\psi$ is the vertical component of the electric field. Note that the wavenumber $k$ and their resonances $\psi$ in the Helmholtz equation play the role of eigenvalues and their wavefunctions, respectively.

In figure~\ref{Fig.1}, the eigenvalues (i.e., wavenumbers) of two interacting modes (denoted by red dashed and blue solid lines) and the intensities of representative wavefunctions [(a), (b), (c), (d), (e), and (f)], as a function of the deformation parameter $\varepsilon$, are plotted in the restricted range $0.75\leq\varepsilon\leq0.11$. In this range, the phase space shows soft chaos (the transition region between regular and chaotic system). In particular, eigenstates (a), (d), (c), and (f) are associated with regular invariant torus. The avoided crossing occurs at $\varepsilon\approx 0.091$. We can also observe that the wavefunctions corresponding to (a) and (d) are mixing together at (b) and (e) in the state of the coherent superposition of wavefunctions. Subsequently, they exhibit an exchanging pattern of wavefunctions at (c) and (f). Here, we investigate the features of this avoided crossing by considering wavefunction (de)localisation.

\section{Analysis of the Inverse Participation Ratio} \label{inverse}

\begin{figure*}
\centering
\includegraphics[width=16.0cm]{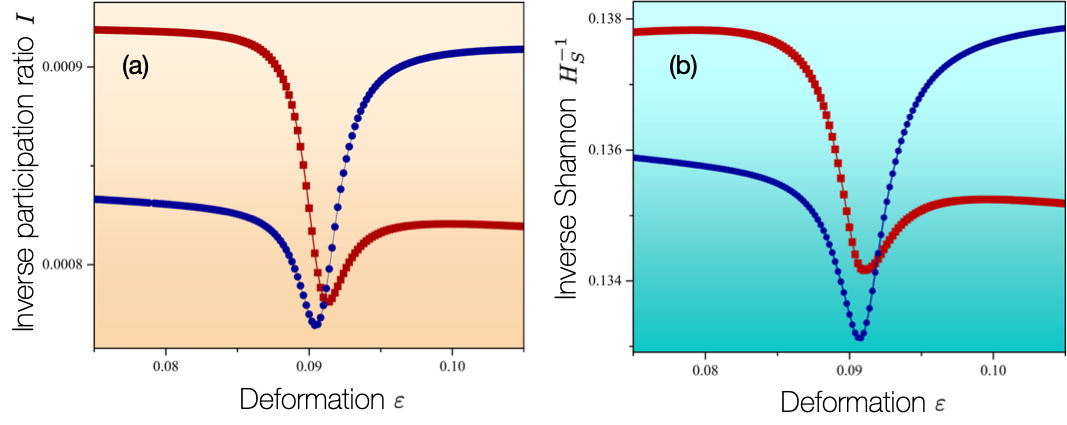}
\caption {(a) Inverse participation ratio ($I$) of intensity distributions of two interacting modes shown in figure.~\ref{Fig.1}. The values of $I$ decrease as the centre of the avoided crossing is approached. (b) Inverse Shannon entropy ($H_{S}^{-1}$) of the intensity distributions of two interacting modes in figure\ref{Fig.1}. The inverse Shannon entropy $H_{S}^{-1}$ in panel (b) presents a trend similar to $I$ in panel (a).}
\label{Fig.2}
\end{figure*}

The inverse participation ratio $I$ is widely employed in percolation systems~\cite{BM17}, flat bands~\cite{S18}, and quantum chaos~\cite{BLR19} to quantify the degree of localisation of the wavefunction with given probability distributions. It is defined by the summation (or integral) over the square of the probability density in the position and momentum space (or phase space). In the case of position space, we can simply define the formula as follows:
 \begin{equation}
I(\rho)=\sum_{j=1}^{N}\rho^{2}(x_{j}),
\end{equation}
where $\rho(x_{j})=|\psi(x_{j})|^{2}$. To apply $I$ to our quadrupole system, we normalise the intensity of the wavefunction as $\sum_{j=1}^{N}|\psi(x_{j})|^{2}=1$. Here, $N$ is the total number of mesh points for the cavity area, and we assume that it corresponds to an effective basis set $\left\{\ket{j}\right\}^{N}_{j=1}$ for the position space. The inverse participation ratio has a maximal value of $I=1$ when the wavefunction is localised at the specific basis state $\ket{j_{p}}$, owing to the fact that $\rho(x_{j})=\delta_{jp}$.

Figure~\ref{Fig.2}(a) presents a plot of the inverse participation ratio of the normalised intensity distributions for the two interacting modes in figure~\ref{Fig.1}. The values of $I$ that are distant from the centre of the avoided crossing have relatively higher values than those at the centre of the avoided crossing. In other words, the values of $I$ decrease when the centre of the avoided crossing is approached. This directly indicates that the wavefunction localisation of two interacting modes increases when the mixing of the wavefunction decreases (i.e., moving away from the centre of avoided crossing).

The inverse Shannon entropy $H_{S}^{-1}$ of the intensity distributions of the two interacting modes (in figure\ref{Fig.1}) is shown in figure\ref{Fig.2}(b) for a relative comparison. The Shannon entropy was obtained in a similar manner as in our previous studies~\cite{PMSK+18,JPK20}, that is, $H_{S}(\{\rho(x_{i})\})= -\sum_{j=1}^{N}\rho(x_{j})\log\rho(x_{j})$. Note, that the Shannon entropy is a relevant measure of the average amount of information content, and we used it as a measure of the degree of uniformity for a given mode pattern. However, we confirm that the Shannon entropy can also be used to measure the wavefunction (de)localisation of the mode pattern; $H_{S}^{-1}$ in panel (b) displays a similar trend with $I$ in panel (a) in figure\ref{Fig.2}.

\section{Quantifying the Delocalisation from R\'{e}nyi Entropy} \label{Renyi}

\begin{figure*}
\centering
\includegraphics[width=16.0cm]{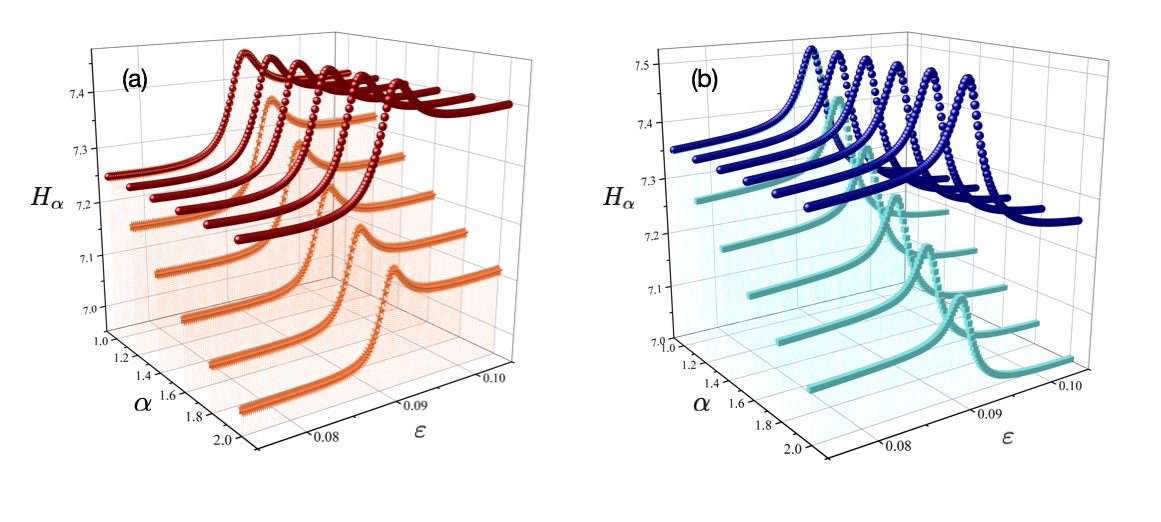}
\caption {(a) Red spheres and orange stars denote the Shannon entropies ($H_{S}$) and R\'{e}nyi entropies ($H_{\alpha}$), respectively. Both are obtained from the intensity distributions of the red mode in figure\ref{Fig.1}. The profiles of $H_{\alpha}$ converge to $H_{S}$ as $\alpha$ approaches $1$. Entropies $H_{S}$ and $H_{\alpha}$ in panel (b) of the blue mode behave similar to those in panel (a).}
\label{Fig.3}
\end{figure*}

Intuitively, the inverse of the inverse participation ratio $I^{-1}$ can be a natural measure of delocalisation. However, we consider the logarithm of $I^{-1}$ instead of $I^{-1}$, which enables us to extend the concept of delocalisation to the information theory.

To achieve this, we first introduce the definition of R\'{e}nyi entropy~\cite{R61,D19}, which is a generalisation of the Shannon entropy, Hartley entropy, min-entropy, and collision entropy~\cite{R61}. The definition of order $\alpha$ R\'{e}nyi entropy for a discrete probability distribution $\{\rho_{j}\}$ (with $\alpha \geq 0$ and $\alpha \neq 1$) is given by
\begin{equation}
H_{\alpha}(\{\rho_{j}\})=\frac{1}{1-\alpha}\log\left(\sum_{j=1}^{N}\rho^{\alpha}_{j}\right),
\end{equation}
under a normalisation condition such that $\sum_{j=1}^{N}\rho_{j}=1$. Note that, the logarithm of $I^{-1}$ is equal to the R\'{e}nyi entropy with $\alpha=2$; that is, $H_{2}(\{\rho_{j}\})\equiv\log(I^{-1})$. In this manner, the delocalisation measured by the inverse of the inverse participation ratio is related to the concept of entropy. Furthermore, the limiting case of $H_{\alpha}$, as $\alpha\to1$, is the Shannon entropy in the form of
 \begin{equation} \label{Shannon}
 \lim_{\alpha\rightarrow1}H_{\alpha}(\{\rho_{j}\})=-\sum_{j=1}^{N}\rho_{j}\log\rho_{j}=H_{S}(\{\rho_{j}\}).
 \end{equation}
 It can be easily proven using the L'H\^{o}spital rule as follows:
 $\lim_{\alpha\rightarrow1}H_{\alpha}(\{\rho_{j}\})=\lim_{\alpha\rightarrow1}\frac{f‘(\alpha)}{g’(\alpha)}=\lim_{\alpha\rightarrow1}\frac{f'(\alpha)}{g'(\alpha)}$, where $f(\alpha)=\log\big(\sum_{j=1}^{N}\rho^{\alpha}_{j}\big)$ and $g(\alpha)=1-\alpha$, and their derivatives $f'(\alpha)$ and $g'(\alpha)$, respectively.

We make use of the R\'{e}nyi entropy to study the wavefunction (de)localisation for avoided crossing in figure\ref{Fig.1}; the explicit results are shown in figure\ref{Fig.3}. As shown in figure\ref{Fig.3}(a), the red spheres and orange stars correspond to the cases of Shannon entropy ($H_{S}$) and R\'{e}nyi entropy ($H_{\alpha}$), respectively, for $\alpha$=2.0, 1.8, 1.6, 1.4, 1.2, and 1.001. Both can be obtained from the normalised intensity distributions of the red mode in figure\ref{Fig.1}, and these are maximised at the centre of the avoided crossing point (i.e., $\varepsilon\approx 0.91$) for each $\alpha$. The $H_{\alpha}$ profiles converge to $H_{S}$ as $\alpha$ approaches $1$, which confirms Eq.~(\ref{Shannon}). In addition, $H_{S}$ and $H_{\alpha}$ in panel (b) were observed to behave similar to those in panel (a) in figure\ref{Fig.3}.

Moreover, when $\alpha$ does not converge to $1$ (the Shannon entropy), particularly when we compare the R\'{e}nyi entropy at $\alpha=2$ with the Shannon entropy, we can easily observe that their profiles are significantly similar to each other, despite their absolute values being different. This fact reveals that Shannon entropy can also be a useful method to quantify the wavefunction (de)localisation for avoided crossing in a quantum billiard.

\section{Root-Mean-Square Image Contrast} \label{rms}

\begin{figure}
\centering
\includegraphics[width=8.0cm]{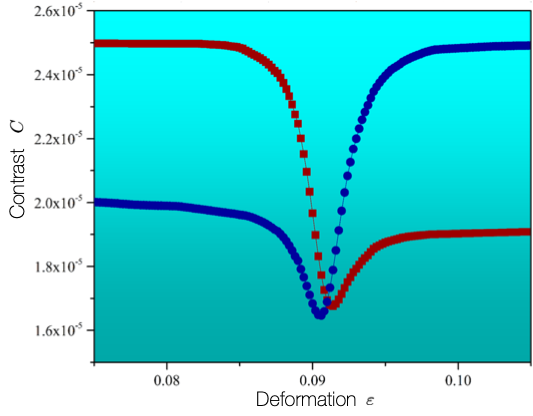}
\caption {RMS image contrast ($C$) of the pixel intensity for two interacting modes in figure\ref{Fig.1}. The two values plotted in red squares and blue circles decrease while the centre of the avoided crossing is approached.}
\label{Fig.4}
\end{figure}

In this section, we assume that the effective basis $\ket{i}$ for the position space corresponds to the pixel of the morphology (or image) of the wavefunction. Then, we can exploit the wavefunction intensity $\rho(x_{j})$ as the pixel intensity $I_{P}(x_{j})$ of the image of the wavefunction. In doing so, the measure of wavefunction (de)localisation can be converted into the measure of the intensity contrast of the image.

For this approach,  we introduce the RMS contrast $C$~\cite{SP94,P90,P87} in terms of the pixel intensity defined as follows:
\begin{equation}
C(I_{P})=\sqrt{\frac{1}{N}\sum_{j=1}^{N}\left(\frac{I_{P}^{j}-\langle I_{P}\rangle}{\langle I_{P}\rangle}\right)^{2}},
\end{equation}
where $\langle I_{P}\rangle$ is the average of all pixel intensities in the image, $I_{P}^{j}$ is the $j$th pixel intensity, and $N$ is the total number of pixels. Contrary to other contrast measures, the RMS contrast does not depend on the specific morphology of the image. That is, we can quantitatively compare the contrast having different morphologies by using RMS contrast~\cite{SP94,P90,P87}. Figure~\ref{Fig.4} presents the results.

The RMS contrast of the pixel intensity of the wavefunctions for two interacting modes in figure~\ref{Fig.1} is shown in figure~\ref{Fig.4}. The two values (indicated by red squares and blue circles) of $C$ decrease as the centre of the avoided crossing is approached, and they are exchanged across the avoided crossing. This behaviour is also significantly similar to the inverse participation ratio in figure~\ref{Fig.2}(a). Thus, the RMS contrast $C$, which is more intuitive and practical, can be used as a measure of the wavefunction (de)localisation property.

\section{Comparison of Three-Type Measures} \label{comparison}

\begin{figure*}
\centering
\includegraphics[width=16.5cm]{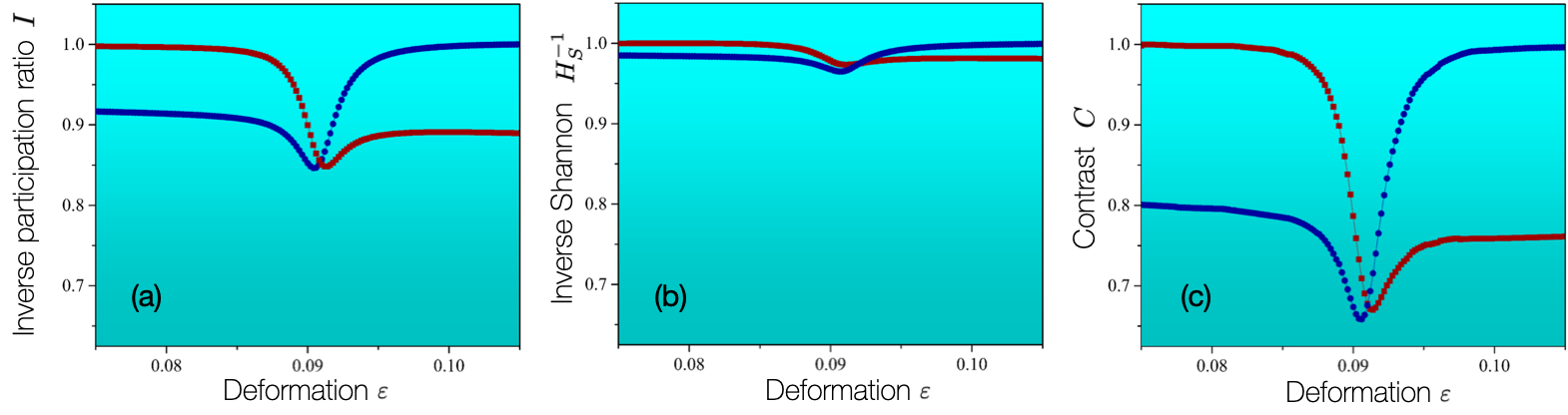}
\caption {Comparison between three types of (de)localisation measures. (a) Normalised inverse participation ratio. (b) Normalised inverse of the Shannon entropy. (c) Normalised RMS contrast. The absolute maximum values of $I$, $H_{S}^{-1}$, and $C$ are approximately $0.000921$, $0.1379$, and $0.000025$, respectively.}
\label{Fig.5}
\end{figure*}
The inverse participation ratio ${I}$, inverse of the Shannon entropy $H_{S}^{-1}$, and RMS contrast $C$, are compared. These quantities are shown in figures~\ref{Fig.5}(a), ~\ref{Fig.5}(b), and ~\ref{Fig.5}(c), respectively. For a fair comparison, we consider that all quantities must be normalised to their maximal values (See the caption in figure~\ref{Fig.5}). It is clearly observed that all of them behave in a similar manner. That is, they are minimised at the centre of the avoided crossing and are also exchanged across the centre of the avoided crossing. These results indicate that ${H_{S}^{-1}}$ and $C$ can also be meaningful indicators for wavefunction (de)localisation along with $I$. Furthermore, we observe that the difference between the maximal and minimal values of $C$ is larger than those of the other two quantities. Thus, we can consider that $C$ is the most sensitive measure at least in our conducted numerical study (not in general case).

However, each quantity has its advantages. The inverse participation ratio ${I}$ is well defined in the position space, momentum space, or in the total phase space~\cite{BLR19}. We can extend the wavefunction (de)localisation to the information theory by addressing ${H_{S}^{-1}}$, whereas $C$ may be a useful tool for interpreting the wavefunction (de)localisation in the framework of image processing tasks. In addition, the localisation measures in previous work~\cite{BLR19} are related to those in this paper but are different. Those in the work~\cite{BLR19} are defined on the phase space and are suitable for analysing the chaotic eigenfunctions whereas our measures are defined on the position space and are more suitable for analysing the effect of the mixing of wavefunctions under the avoided crossings.

\begin{figure*}
\centering
\includegraphics[width=13.5cm]{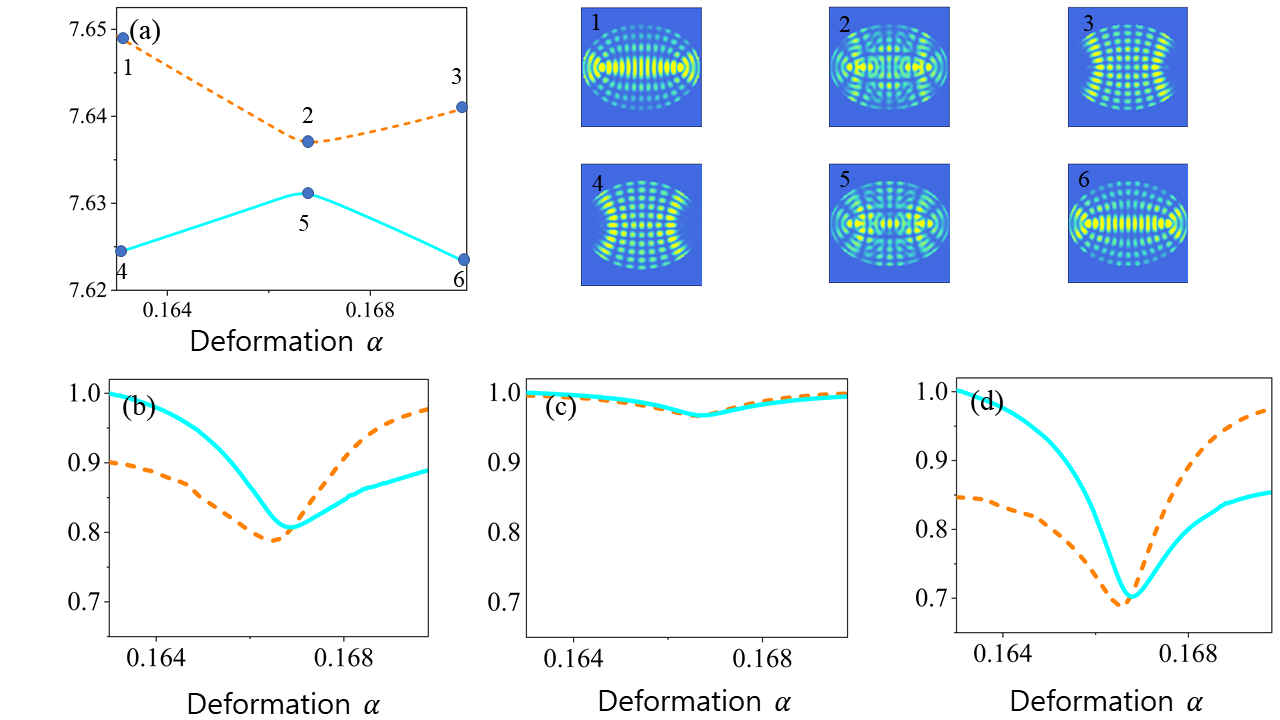}
\caption {(a) Eigenvalues (or wavenumbers) of two interacting eigenmodes near the centre of the avoided crossing and the intensities of representative wavefunctions ($1,2,3,..6$) for an open elliptic billiard as a function of deformation parameter $\alpha$ at $n=3.3$. The deformation parameter $\alpha$ is associated with with the major axis $a = R(1 +\alpha)$ and the minor axis $b=(\frac{R}{1 +\alpha})$. (b) Normalised inverse participation ratio. (c) Normalised inverse of the Shannon entropy. (d) Normalised RMS contrast}
\label{Fig.6}
\end{figure*}

In addition, to demonstrate the general validity of our argument, we apply the three-type measures to an open elliptic billiard in ~\ref{Fig.6}. We notice that a similar trend is observed with the closed quadrupole billiard. This is not a trivial result sine the avoided crossing in ~\ref{Fig.1} comes from the nonlinear dynamical effects in a chaotic system whereas the avoided crossing in ~\ref{Fig.6} comes from the openness effects in non-Hermitian system~\cite{PMSK+18,PJ18}.
\section{Conclusions} \label{conclusion}
The inverse participation ratio, R\'{e}nyi entropy, and RMS image contrast were investigated in a quadrupole billiard and an open elliptic billiard for the measurement of the wavefunction (de)localisation in an avoided crossing. The inverse participation ratio, as a measure of wavefunction localisation, is minimised at the centre of the avoided crossing and increases towards the ends of the avoided crossing. For the extension from the wavefunction delocalisation to the information theory, we employed the logarithm of the inverse of the inverse participation ratio instead of the inverse of the inverse participation ratio.

We confirmed that Shannon entropy can also be another indicator of wavefunction delocalisation. Moreover, we exploited the RMS image contrast of the pixel intensity of the mode pattern to quantify the wavefunction localisation, and the RMS image contrast was minimised as the centre of interaction approached. Consequently, we confirmed that the strength of the interaction can be related to the image contrast. The three indicators of wavefunction (de)localisation behave similarly; however, the RMS image contrast is the most sensitive at least in our presented example. The eigenvalues that we have used in this paper are in the low-energy region, and not in the semiclassical region. Apparently, one can expect that the effects of (de)localisation would be enhanced in the semi-classical (high-energy) range. We will address these conjectures on phase space as well as configuration space in our future works.
\section*{Acknowledgments}
We thank Kyungwon An for valuable comments. This work was supported by the National Research Foundation of Korea, a grant funded by the Ministry of Science and ICT (Grant No. NRF-2020M3E4A1077861 \& 2020R1A2C3009299) and the Ministry of Education (Grant No. NRF-2021R1I1A1A01052331 \& NRF-2021R1I1A1A01042199).

\end{document}